# Synthesis of Bulk Superconducting $LiNbO_2$ Crystals through $CaH_2$ Reduction

*Ryan Paxson[1,2,*], Stephanie J. Hong[3], Bicky S. Moirangthem[6], Parham Kabirifar[5], Saya Takeuchi[1], Tianyu Li[3], Chih-Yu Lee[1], Haotong Liang[1], Keenan Avers[2], Kamal R. Joshi[6], Amlan Datta[6], Makariy A. Tanatar[6], Shanta Saha[2], Peter Zavalij[3], Ruslan Prozorov[6], Alexander J. Grutter[4], Efrain E. Rodriguez[1,2,3,*], and Ichiro Takeuchi[1,2,*]*

[1]*Department of Materials Science and Engineering, University of Maryland, College Park, MD, USA*

[2]*Maryland Quantum Materials Center, University of Maryland, College Park, MD, USA*

[3]*Department of Chemistry & Biochemistry, University of Maryland, College Park, MD, USA*

[4]*NIST Center for Neutron Research, National Institute of Standards and Technology, Gaithersburg, MD, USA*

[5]*Advanced Imaging and Microscopy (AIM) Lab, Maryland Nanocenter, University of Maryland, College Park, MD, USA*

[6]*Ames Laboratory and Department of Physics and Astronomy, Iowa State University, Ames, IA, USA*

*We have synthesized layered superconducting $LiNbO_2$ crystals through a bulk phase transformation from $LiNbO_3$ single crystals via $CaH_2$ reduction. As the Nb valence is reduced from 5+ to 3+, the material undergoes a structural transformation to the resulting product, $LiNbO_2$, which is accompanied by metallic behavior and a superconducting transition, $T_c$ onset, as high as 14.4 K. Secondary ion mass spectroscopy (SIMS) and X-ray photoelectron spectroscopy (XPS) show that the resulting phase is hole-doped through de-lithiation during the reduction. Magnetization and AC susceptibility measurements from a tunnel diode resonator confirm the bulk nature of superconductivity with a superconducting volume fraction of ≈ 77% and an upper critical field approaching 26 T. Our study demonstrates extreme hydride reduction as an effective method to induce phase transformations with non-topotactic pathways and can be used to synthesize bulk materials with exotic properties.*

## 1. Introduction

Hydride assisted reduction has proven to be an effective method to tune the properties of metal oxides.[1,2] Through anion stoichiometry manipulation, e.g. inserting $H^-$ ions and removing oxygen ions from the lattice, hydride assisted reduction gives rise to myriads of interesting phenomena. This includes structural transformations, new magnetic structures, and the stabilization of rare oxidation states (e.g. $Co^+$ and $Ni^+$).[3,9,10,12-15,17-19] This process differs from conventional gas reduction methods (e.g. via the use of $H_2$ gas) because the reaction requires a much lower temperature, which assists the formation of metastable oxyhydrides and oxygen deficient compounds.[3-12,22,28,29] Calcium hydride ($CaH_2$) reduction has been used to synthesize the infinite-layer square planar lattice in nickelates, which hosts robust superconductivity.[7,10,11,16-19] While the discovery of superconductivity in infinite-layer nickelates has fueled an explosion of interest in the $CaH_2$ reduction process and its ability to access superconducting phases in perovskite metal oxides, the overwhelming majority of applications are focused on topotactic processes in thin film samples with sub-nL volumes. The extent to which $CaH_2$ reduction can be used to synthesize bulk superconductors through non-topotactic transformations or *via* the reduction of large-scale single crystals remains relatively unexplored. In this work, we demonstrate $CaH_2$ reduction of $LiNbO_3$ single crystals to produce a bulk volume of superconducting $LiNbO_2$.

The oxygen-rich parent compound, $LiNbO_3$, orders in a distorted trigonal structure with Nb cations coordinated by 6 oxygen atoms in an octahedral environment as shown in **Figure 1**(a).[34] Unlike the perovskite to square-planar transformation of the nickelate superconductor, the transformation of $LiNbO_3$ into superconducting $LiNbO_2$ cannot proceed *via* a topotactic process due to the different cation sublattice of the layered $P6_3/mmc$ space group of $LiNbO_2$, shown in Figure 1(b).[24-27] Instead, the cation sublattice structure changes completely during the phase transformation. Thus, unlike the perovskite to infinite-layer square lattice case, the $CaH_2$ reduction facilitates a reconstructive phase transformation from lithium niobate to lithium niobite requiring a complete reorganization of the atoms in the lattice.

$LiNbO_2$ has been shown to be superconducting with either hole or electron doping, and is commonly hole-doped through delithiation or electron doped by hydrogen incorporation .[24-27] Hydrogen incorporated either as a cation on Li sites or an anion on oxygen sites both contribute to electron doping the sample, and counteract the hole doping from delithiation.[20,26] Both hole and electron doping in $LiNbO_2$ are compensated by changing the oxidation state of the Nb atoms.

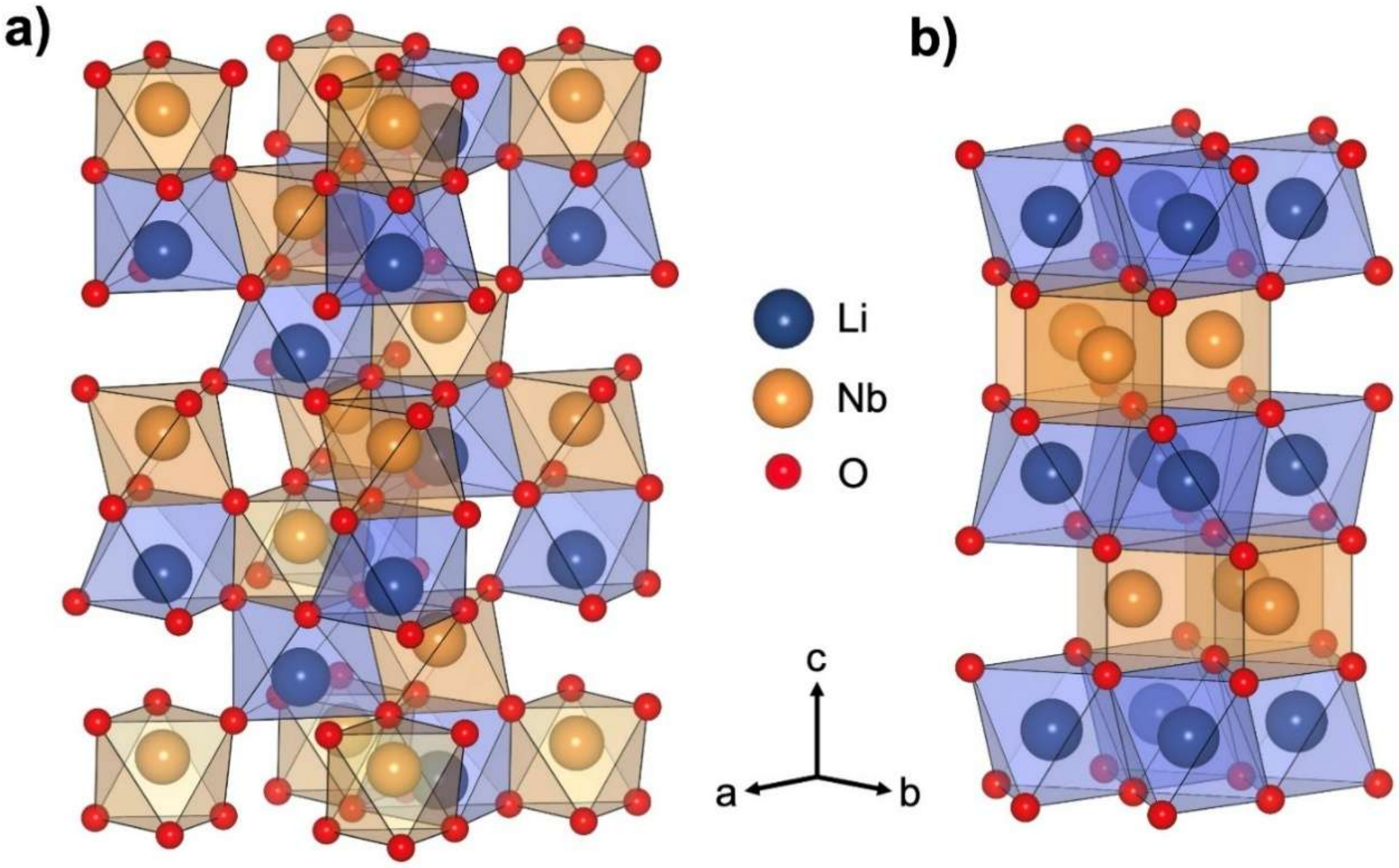

**Figure 1.** Crystal structures of (a) $LiNbO_3$ and (b) $LiNbO_2$ with the $CsPrS_2$ structure type.

While past theoretical and experimental work suggest a wealth of exciting physics in $LiNbO_2$, including possible unconventional two-dimensional superconductivity in $NbO_2$ layers mediated through strong electron correlations, detailed investigations have been inhibited by the lack of bulk crystal samples.[35,36] Thus far, superconductivity in $LiNbO_2$ has only been reported in powders and thin films.[24-27] In this work, we demonstrate the utility of $CaH_2$ reduction to induce bulk superconductivity at scale, reducing crystals with masses approaching 200 mg. We observe the onset temperature, $T_c^{onset}$, of the superconducting transition and $T_{c0}$, at which the resistance reaches zero, as high as 14.4 K and 13.3 K, respectively. These are comparable to the highest reported superconducting transition values for $LiNbO_2$ reported in the literature. The successful application of $CaH_2$ reduction to macroscopic volumes of a more complex system with a distorted lattice, volatile lithium and a non-topotactic transformation pathway highlights the adaptability and promise of this approach.

## 2. Results

### 2.1. Synthesis of $LiNbO_2$ and structural and composition analysis

To perform the hydride reduction of $LiNbO_3$, single crystal $LiNbO_3$ of varying orientations (Y-cut, Z-cut and Y-128 °) were covered with 0.5 g of $CaH_2$ powder and sealed in evacuated (approximately 0.133 Pa) quartz ampoules. The Y-128° is an oblique crystallographic axis rotated 128° from the c-axis with respect to the a-axis, which exposes a loosely packed and highly corrugated surface compared to the Y and Z-cut crystals. The hydride reduction was conducted at various temperatures (350 °C to 700 °C) and for between 72 hours and 720 hours. During the reduction, the sample undergoes a dramatic color change: the pale-yellow transparent $LiNbO_3$ crystal turns dark black with a metallic luster. The highest $T_c$ onset of 14.4 K was observed for the Y-128° crystal reduced for 240 h at 650 °C temperature (see below).

**Figure 2** shows the scanning electron microscopy cross-sectional images and analysis of the reduced $LiNbO_3$ crystal (Y-128°). Two distinct regions can be observed in the crystal as seen in Figure 2(a).

The pristine $LiNbO_3$ sample, with dimensions 10 mm x 10 mm x 0.5 mm was reduced by the $CaH_2$ reaction from the surfaces, creating bulk layers of $LiNbO_2$ at its top and bottom surface regions, leaving a layer of the unreacted parent compound $LiNbO_3$ in the center. Energy dispersive X-ray spectroscopy (EDS) mapping of oxygen, shown in Figure 2(b), confirms the regions converted from the $CaH_2$ reaction are oxygen reduced, and the EDS line scans shown in Figure 2(c), indicate that the $LiNbO_2$ regions show relatively uniform oxygen stoichiometry, with an approximate ratio of 3 (center bulk region) to 2 (surface bulk regions), further indicative of the fact that the $CaH_2$ reaction led to the phase transformation from $LiNbO_3$ to $LiNbO_2$, rather than merely inducing defects such as oxygen vacancies. The contrast between the reduced and unreduced layers is more easily discernible through an optical cross-sectional microscope image, shown in Figure 2(d). The thickness of the $LiNbO_2$ regions was found to increase with the $CaH_2$ reduction time. To our knowledge, this study is the first to report the conversion of bulk $LiNbO_3$ single crystals to $LiNbO_2$.

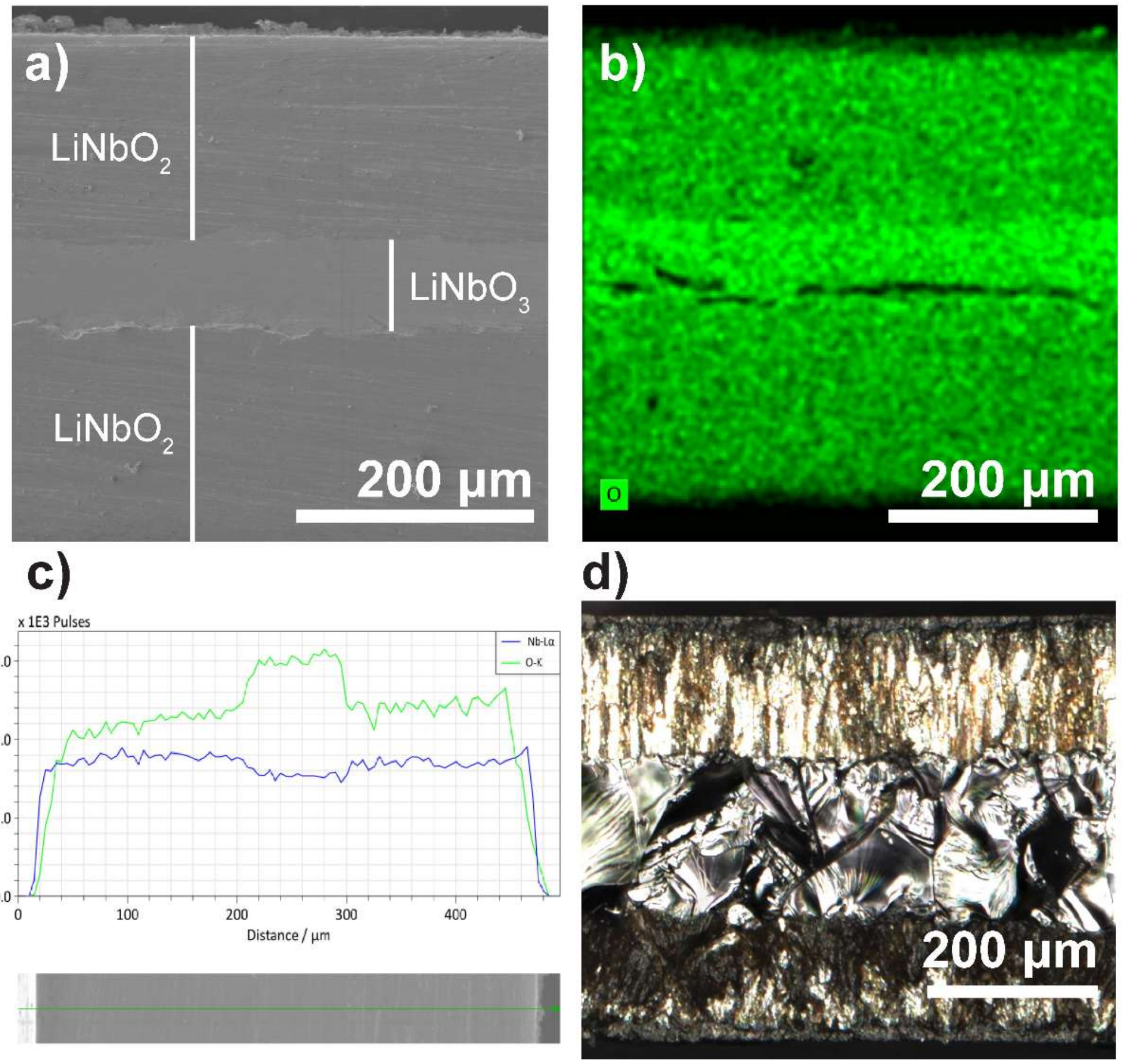

**Figure 2.** (a) Cross-sectional SEM image of $LiNbO_3$ sample which was $CaH_2$ reduced for 480 hours at 650 °C. (b) Oxygen Energy Dispersive X-ray Spectroscopy (EDS) mapping shows the regions formed from the $CaH_2$ reaction are oxygen reduced. (c) EDS line scans, indicate the $LiNbO_2$ regions are uniform in oxygen stoichiometry and gradients in oxygen content are not observed. (d) Optical image of reduced sample cross section using digital microscope.

**Figure 3** shows the powder X-ray diffraction pattern of the $LiNbO_2$ crystal region, which was removed from the surfaces of a naturally formed $LiNbO_2/LiNbO_3/LiNbO_2$ bulk heterostructure and crushed into a polycrystalline form. This sample was reduced from Y-128° oriented $LiNbO_3$ at 650 °C for 480 h. The reduced material contains 80.2(3)% of the layered $P6_3/mmc$ $LiNbO_2$ by weight, while the remaining 19.8(3)% is a remanent $LiNbO_3$ phase. We note that no diffraction peaks associated with metallic Nb or $NbO_x$ phases are observed in this analysis. The non-topotactic nature of the phase transformation from $LiNbO_3$ to $LiNbO_2$ suggests that the resulting crystals need not maintain the same crystallographic orientation as the parent substrate, or even a uniform orientation throughout the reduced sample.

To evaluate the crystallographic orientation, we performed Laue diffractometer measurements and high-resolution X-ray diffraction (HRXRD) scans on reduced Y-128° samples, as is shown in **Figure 4**. The presence of localized, broad (00L) $LiNbO_2$ Bragg reflections in the 2D HRXRD scan in Figure 4(a) suggests a preferred orientation in the reduced samples. The broadened peaks at 17° and 34.2° are attributed to the (002) and (004) reflections of $LiNbO_2$, respectively. The broadening of the (00L) reflections in $\chi$, the horizontal tilt axis, is consistent with fiber texture or mosaic spread in $LiNbO_2$. This broadening is supported by the optical microscope images in Figure 2(d), where vertical columnar crystalline grains in the top $LiNbO_2$ layer can be seen with the typical size of 9 µm. Interestingly, the $LiNbO_2$ phase is oriented with the c-axis normal to the surface of the sample, unlike the parent compound which has its crystalline axis offset from the surface by 38°. The remaining peaks show rings in $\chi$, consistent with the presence of randomly oriented grains. The peaks at 32.3°, 34.9°, and 38.6° are attributed to the (104), (110) and (006) reflections of the parent compound $LiNbO_3$ respectively. Weakly diffracting rings which are not indexed are assumed to be impurities in the overly reduced sample surface, as no significant impurity phases are seen in the powder XRD scan shown in Figure 3.[23] The Laue diffraction pattern in Figure 4(c) shows the coexistence of polycrystalline rings as well as broad localized reflections. This suggests a mix of "powder-like" polycrystallinity which also includes significant clusters with some preferred (00L) orientation normal to the surface.

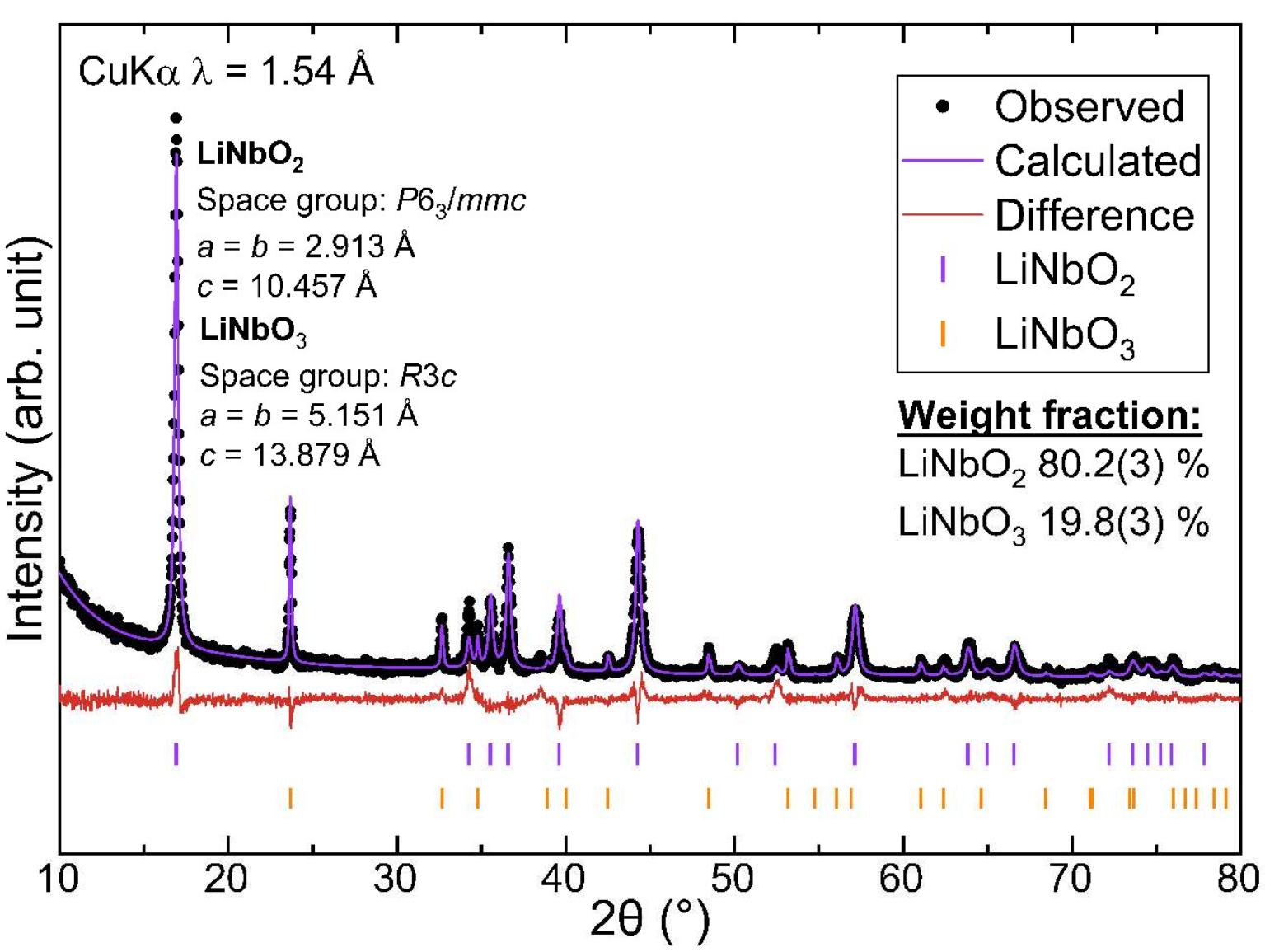

**Figure 3.** Rietveld refinement with powder XRD (CuKα λ = 1.54 Å) data of $LiNbO_3$ sample which was $CaH_2$ reduced for 480 h at 650 °C.

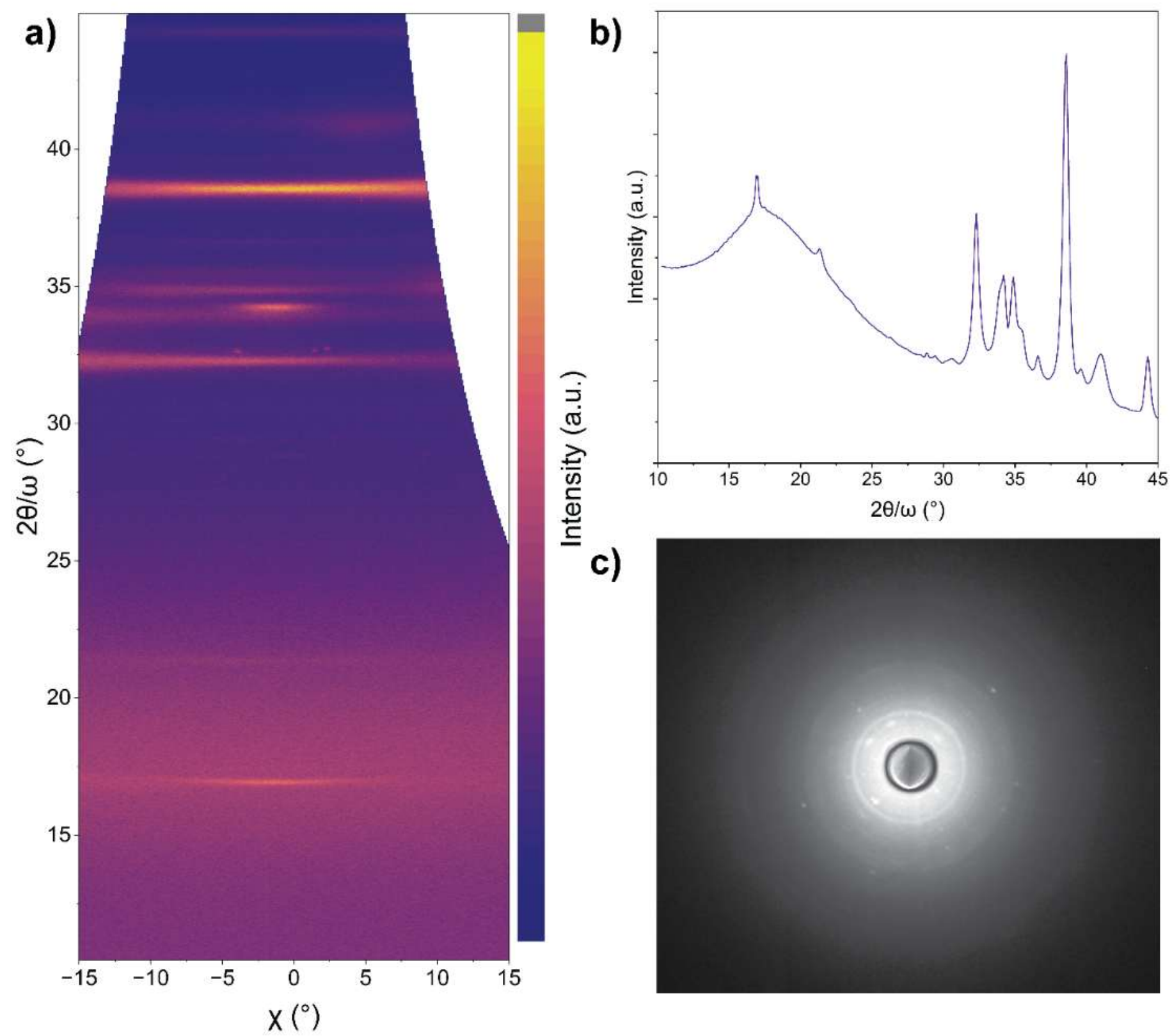

**Figure 4.** Two-dimensional HRXRD scan (a), showing structural anisotropy of reduced sample. The 2D HRXRD scan is integrated and plotted in one dimension shown in (b). Laue diffraction image of reduced sample (c) showing some Bragg reflection along with diffuse scattering.

Quantitatively determining the lithium and hydrogen stoichiometry of $LiNbO_2$ is challenging, so to investigate qualitative changes in lithium and hydrogen stoichiometry after $CaH_2$ reduction, secondary ion mass spectroscopy (SIMS) was performed on the as-reduced $LiNbO_2$ "surface" region as well as an unreduced $LiNbO_3$ crystal (as a control) (**Figure 5**). For the analysis, the lithium and hydrogen species were normalized to niobium oxide ions. This assumption is supported by the cross-sectional EDS line scan shown in Figure 1(c). After reduction, SIMS shows that both delithiation and hydrogenation have occured during the process. The $Li_2^+/NbO^+$ ratio for the $LiNbO_2$ sample after reduction is almost an order of magnitude lower than the control, which is shown in Figure 5(a). The SIMS measurements also reveal gradients in lithium stoichiometry as a function of depth into the sample, with a very thin layer near the surface exhibiting a significantly higher $Li_2^+$ intensity compared to deeper in the crystal. We speculate that lithium migrates to the surface during the $CaH_2$ reduction and escapes the sample. However, SIMS is only suitable for characterizing the upper regions of the crystal and primarily provides information near the surface. Understanding the Li stoichiometry deeper in the sample requires additional study. Hydrogenation has also likely occurred in the sample as the reduced $LiNbO_2$ samples show almost an order of magnitude increase in the $H^-/NbO_2^-$ ratio (Figure 5(b)). Since there is presumed to be no excess hydrogen in the pristine $LiNbO_3$ control sample, the $H^-/NbO_2^-$ ratio of the control sample serves as background hydrogen concentration.

To further understand the role of doping in our $LiNbO_2$ samples, X-ray photoelectron spectroscopy (XPS) is carried out before and after the $CaH_2$ reduction process at 650 °C for 240 h. **Figure 6** (a) and (b) show the XPS spectra of Nb 3*d* and Nb 4*s*/Li 1*s*, respectively, for the $LiNbO_3$ sample before the $CaH_2$ reduction process. The positions of both $Nb^{5+}$ 3d 5/2 and $Nb^{5+}$ 3*d* 3/2 agree with those of the reported spectra on $LiNbO_3$.[30] XPS measurements were taken after polishing the reduced $LiNbO_2$ surface to determine the lithium stoichiometry in the bulk of the sample. Figure 6(c) and (d) show the XPS spectra of the Nb 3*d* and Nb 4*s*/Li 1*s*. The Nb 4*s* and Nb 3*d* spectra in the reduced sample show significant deviations from those of the pristine $LiNbO_3$ single crystal: most notable is the presence of large $Nb^{3+}$ peaks associated with the oxygen removal. In the reduced sample, the Nb spectra are asymmetric as compared to the raw $LiNbO_3$ substrate, reflecting the metallic behavior of the reduced sample. Both Nb 4*s* and Nb 3*d* regions show some overlaps with the Nb spectra of $LiNbO_3$ raw substrate, indicating the existence of residual $Nb^{5+}$ even after the hydride treatment. We fit the Nb spectra as a mixture of the $Nb^{5+}$ and $Nb^{3+}$ oxidation states. The $Nb^{3+}$ oxidation state has its 3*d* 5/2 and 3*d* 3/2 peaks positioned at 203 eV and 206 eV, respectively. The ratio of $Nb^{5+}$ and $Nb^{3+}$ gives an average oxidation state of +3.86.[32,33] Thus, the overall XPS spectra suggests the $LiNbO_2$ regions in the crystal is hole doped. Interestingly, the elemental ratio calculated from XPS in the bulk of the $LiNbO_2$ sample suggests a stoichiometry of $Li_{0.96\pm0.225}NbO_{1.83\pm0.187}$, with a range of measured lithium values from $Li_{0.61}$-$Li_{1.13}$. It is worth noting that the Li 1*s* spectra overlaps with Nb 4s spectra for the treated $LiNbO_3$ crystal, which can lead to large uncertainties in the Li/Nb ratio. The SIMS depth profiles of the reduced sample indicated large lithium pileup and gradients as a function of depth into the sample and a region of very high lithium concentration near the surface. XPS measurements of the as-reduced surface of the $LiNbO_2$ sample are in excellent agreement with the SIMS, and indicate a region at the surface with very high lithium content as shown and discussed in Figure S1. O 1s spectra for all measured samples are shown in Figure S2**.** Based on the XPS spectra and SIMS depth profile, we can conclude that likely both delithiation and $H^-$ insertion is occurring in our samples acting as hole and electron dopants respectively. The absence of niobium metal peaks in the XPS spectra as well as the powder XRD data further supports our conclusion that no metallic niobium clusters are formed at the surface during the hydride treatment process.

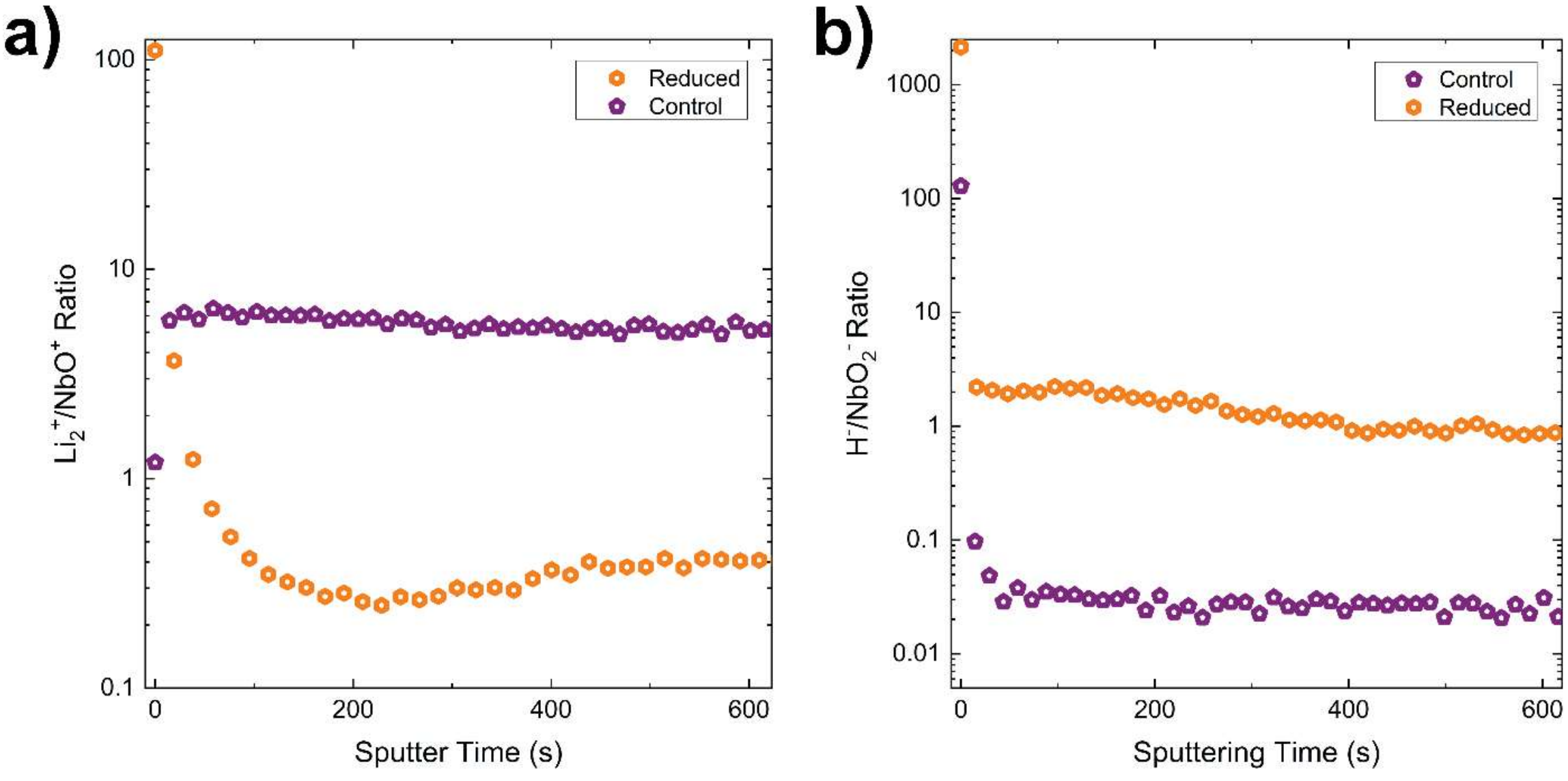

**Figure 5.** Semi log plots of SIMS depth profiles of an unreduced, control $LiNbO_3$ sample and a reduced $LiNbO_2$ sample. (a) Lithium stoichiometry is compared through changes in the $Li_2^+/NbO^+$ ratio. (b) Differences in hydrogen concentration after reduction is evident by looking at the $H^-/NbO_2^-$ ratio between both samples.

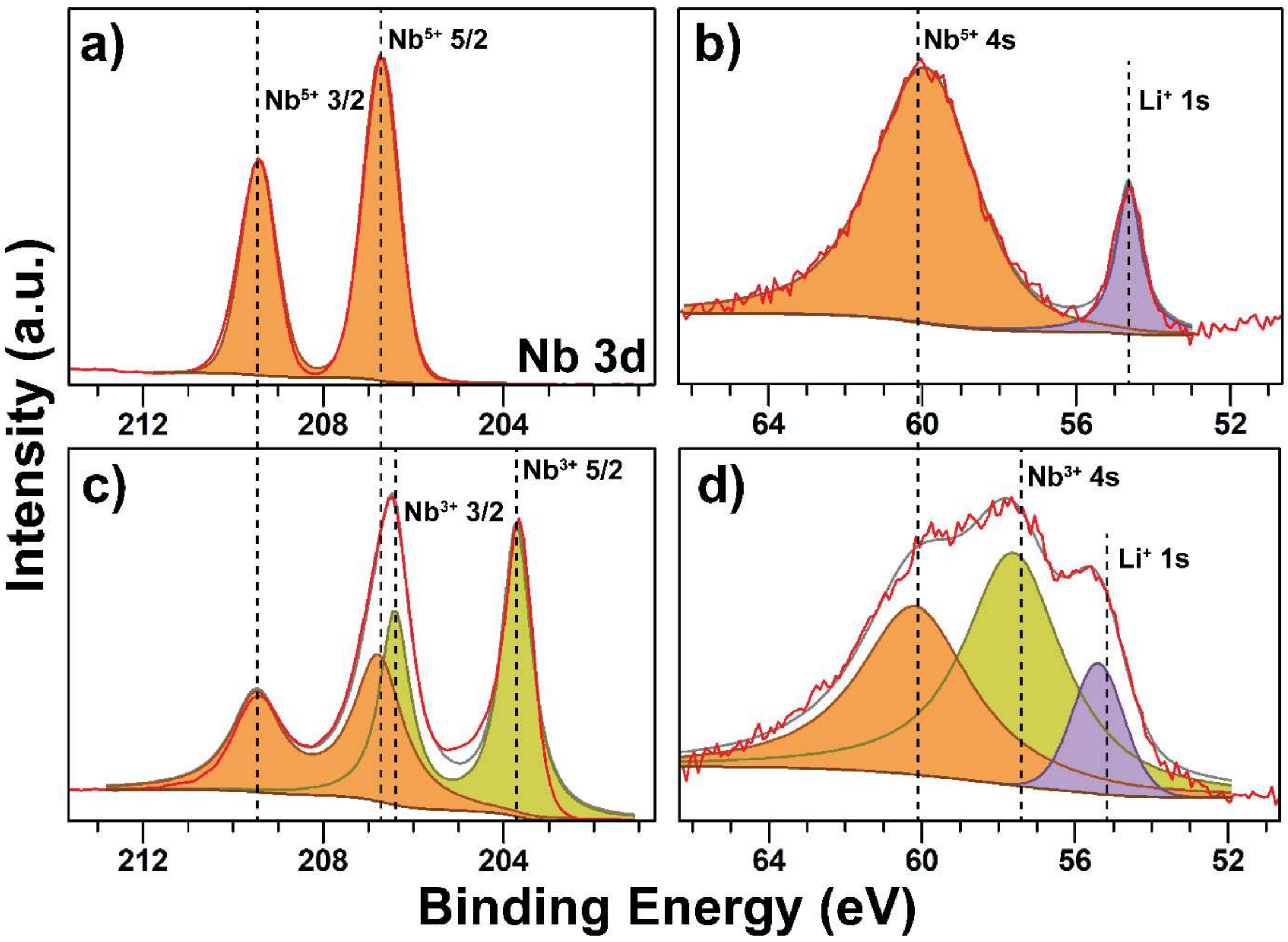

**Figure 6.** XPS Spectra of control $LiNbO_3$ (a-b) and polished $LiNbO_2$ (c-d) samples. Nb 3*d* spectra and fittings are shown in (a) and (c). Li 1*s* and Nb 4*s* spectra and fittings are shown in (b) and (d).

### 2.2 Superconducting Properties

We now turn to the superconducting properties of reduced $LiNbO_2$ crystals. In **Figure 7**(a), a Y-128° oriented $LiNbO_3$ sample subjected to the $CaH_2$ reduction process at 650 °C for 120 h, has transformed into metallic $LiNbO_2$ with a resistivity of approximately 0.45 mΩ cm at room temperature. A superconducting transition is observed in this $LiNbO_2$, with the onset temperature, $T_c^{onset}$, of approximately 14.4 K, and a zero-resistance temperature of 13.3 K. The superconductivity of $CaH_2$ reduced nickelates are reported to emerge when the reduction is performed at around 300 °C.[11] For bulk $LiNbO_3$, however, 650 °C seems to be the ideal reduction temperature for optimizing reaction time and critical temperature.
We examine the dependence of the superconducting transition on the crystal orientation of the starting $LiNbO_3$ single crystals (10 mm x 10 mm x 0.5 mm), namely Y-cut, Z-cut and Y-128°. As shown in Figure 7(c), superconducting critical temperatures are maximized at different reaction times depending on the crystal orientation, suggesting diffusion plays a role in the $CaH_2$ reaction and growth of the reduced phase. The miscut Y-128° oriented crystal exhibits the highest $T_c^{onset}$ despite requiring the least amount of reaction time to achieve this. We postulate that during the high temperature reduction, the more loosely packed, stepped terrace surface allows Li and O to diffuse out of the material more easily as compared to the densely packed and atomically flat surfaces of the Y-cut and Z-cut crystals. The highest $T_c^{onset}$ observed here for a $LiNbO_2$ phase of hydride treated $LiNbO_3$ is 14.4 K, which is among the highest compared to previously reported $LiNbO_2$-related superconductors, e.g., delithiated $LiNbO_2$ powder (5.5 K) , directly synthesized $Li_{1-x}NbO_2$ powder (14 K), hydrogen inserted $LiNbO_2$ (5.5 K) and delithiated $LiNbO_2$ epitaxial films (4.2 K).[24-27]
The upper critical field, $H_{c2\perp}$, was determined from the suppression of superconductivity under magnetic fields applied perpendicular to the plane of the crystal. Resistivity vs. temperature curves in different fixed external magnetic fields are shown in Figure 7(b). The extrapolation of the resistivity to its normal state values determines $T_c(H)$. It is plotted in a conventional form as $H_{c2}(T)$ shown in (Figure 9(b)). The solid orange line in is the fitted field response according to an approximate Helfand-Werthamer equation, yielding a $H_{c2}(0)$ of 25.92 T.[37,39]

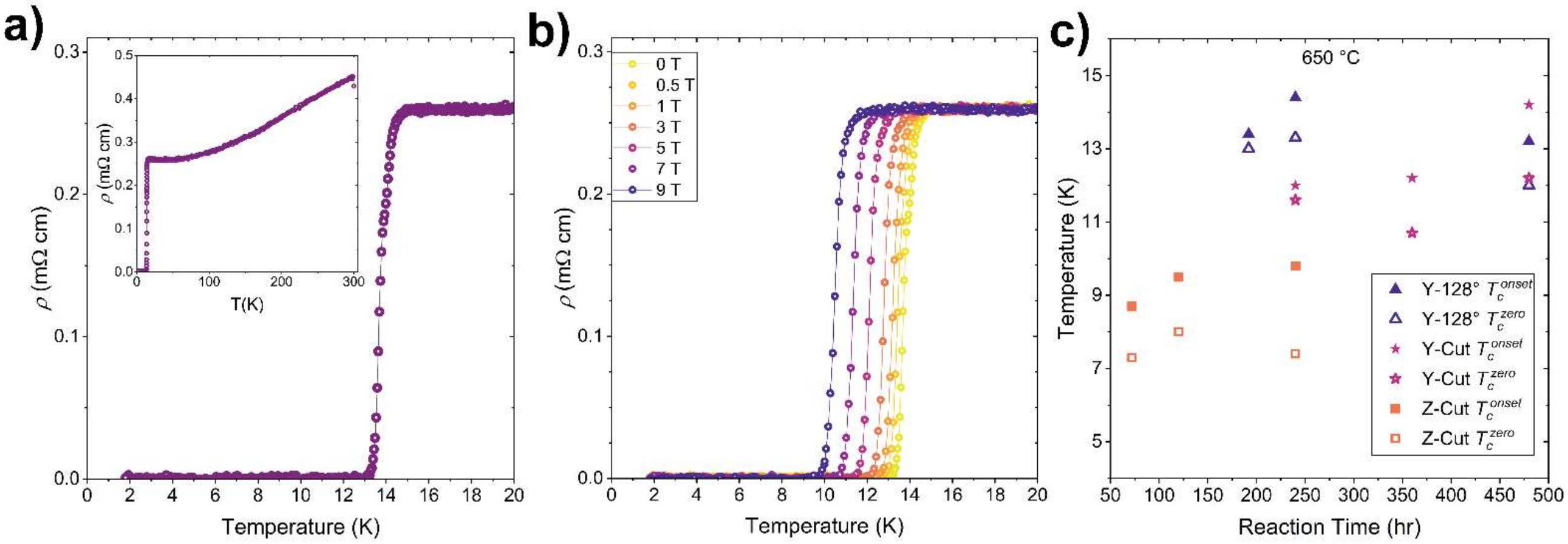

**Figure 7.** a) Resistivity vs. Temperature curve for sample which was reduced for 480 h at 650 °C. b) Resistivity vs. Temperature curves with different external fields applied perpendicular to the direction of the current. c) Critical temperature vs. $CaH_2$ reaction times for substrates with various orientations.

The SQUID measurements show the presence of a robust bulk Meissner effect in the superconducting samples. The zero-field-cooled (1 mT) magnetization vs. temperature (M-T) curve for a $LiNbO_2$ sample (reduced at 650 °C for 480 hours) is shown in **Figure 8**. From the M-T curve the superconducting volume fraction ($V_{sc}$) was calculated using the formula $V_{SC} = 4\pi\chi$, where $H$ is the applied magnetic field strength and $\chi$ is the volume magnetic susceptibility at $T$=1.8 K. The dimensions of the $LiNbO_2$ sample measured are 4 mm x 3 mm x 370 μm (excluding the unreacted $LiNbO_3$ region), where the thickness of the reduced layers was determined from the cross-sectional SEM images shown in Figure 1(a). Taking the demagnetization factor N = 0.86 (calculated using shape of a rectangular cuboid in out-of-plane magnetic field) into consideration, we arrive at the superconducting volume fraction of approximately 77%.[38]

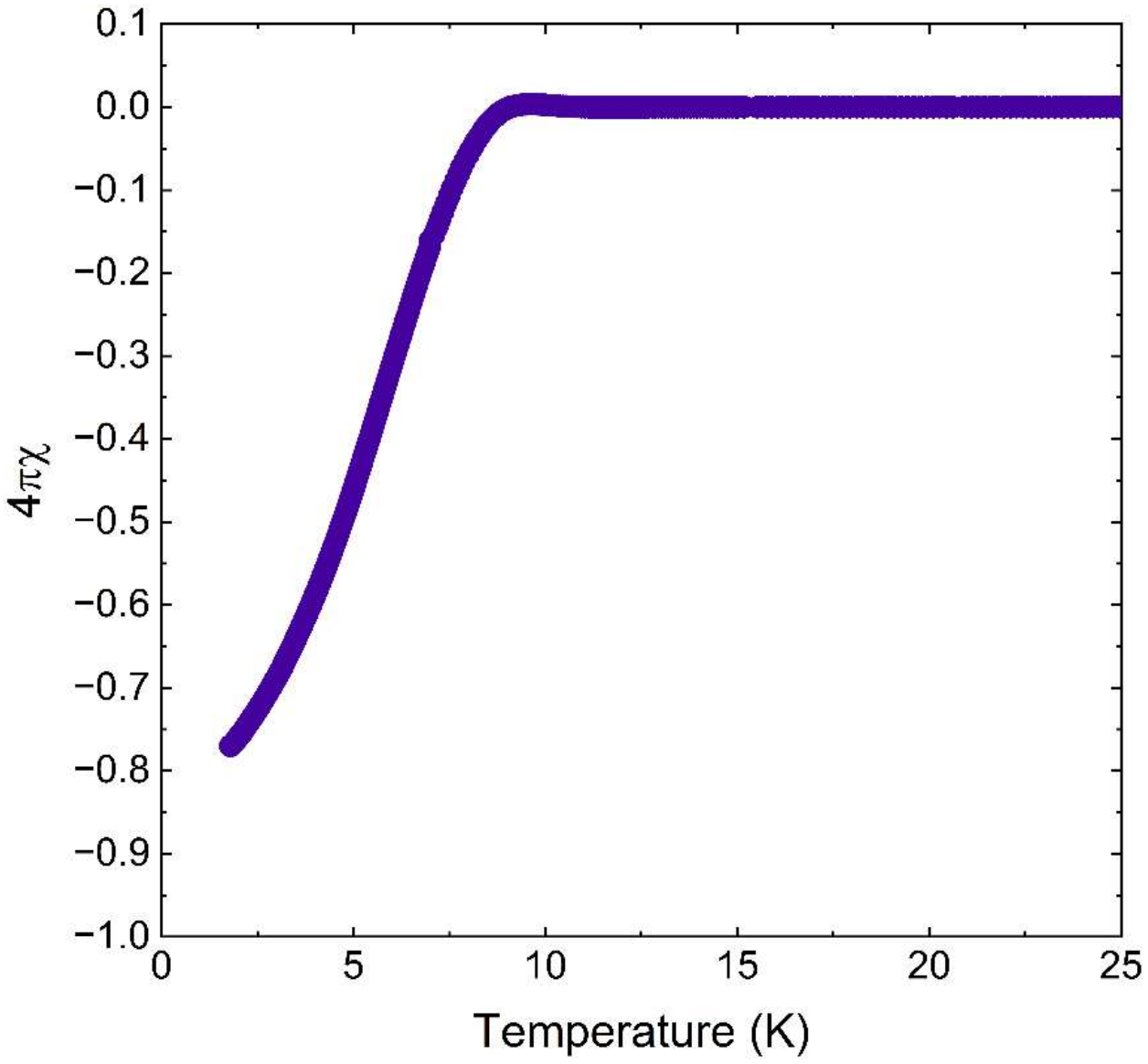

**Figure 8.** Meissner effect shown in ZFC Magnetization vs. Temperature for $LiNbO_2$ sample. Magnetic field of 1 mT is applied out of plane with respect to the sample surface. This sample was synthesized through $CaH_2$ at 650 ºC for 480 h.

Furthermore, the precise Meissner state magnetic response in a crystalline sample with dimensions of approximately 0.3 mm x 0.3 mm x 0.2 mm was measured using a tunnel diode resonator down to 400 mK. The temperature dependent magnetic susceptibility χ(T) during field cooling (FC) in various applied DC magnetic fields is shown in **Figure 9**(a). The low temperature part of the zero-field curve shows a distinct convex downturn which disappears slowly with the increase in field strength. Such behavior in χ suggests the presence of a proximity effect in the neighboring non-reduced layers.[40] The $H_{c2}(T)$ is shown in Figure 9(b). Fitting with an approximate form of the Helfand-Werthamer equation yields $H_{c2}(0)$ of 24.28 T which is in good agreement with the result obtained from our transport measurements.[37,39] Together these measurements confirm the robust, bulk nature of superconductivity in our reduced $LiNbO_2$ samples.

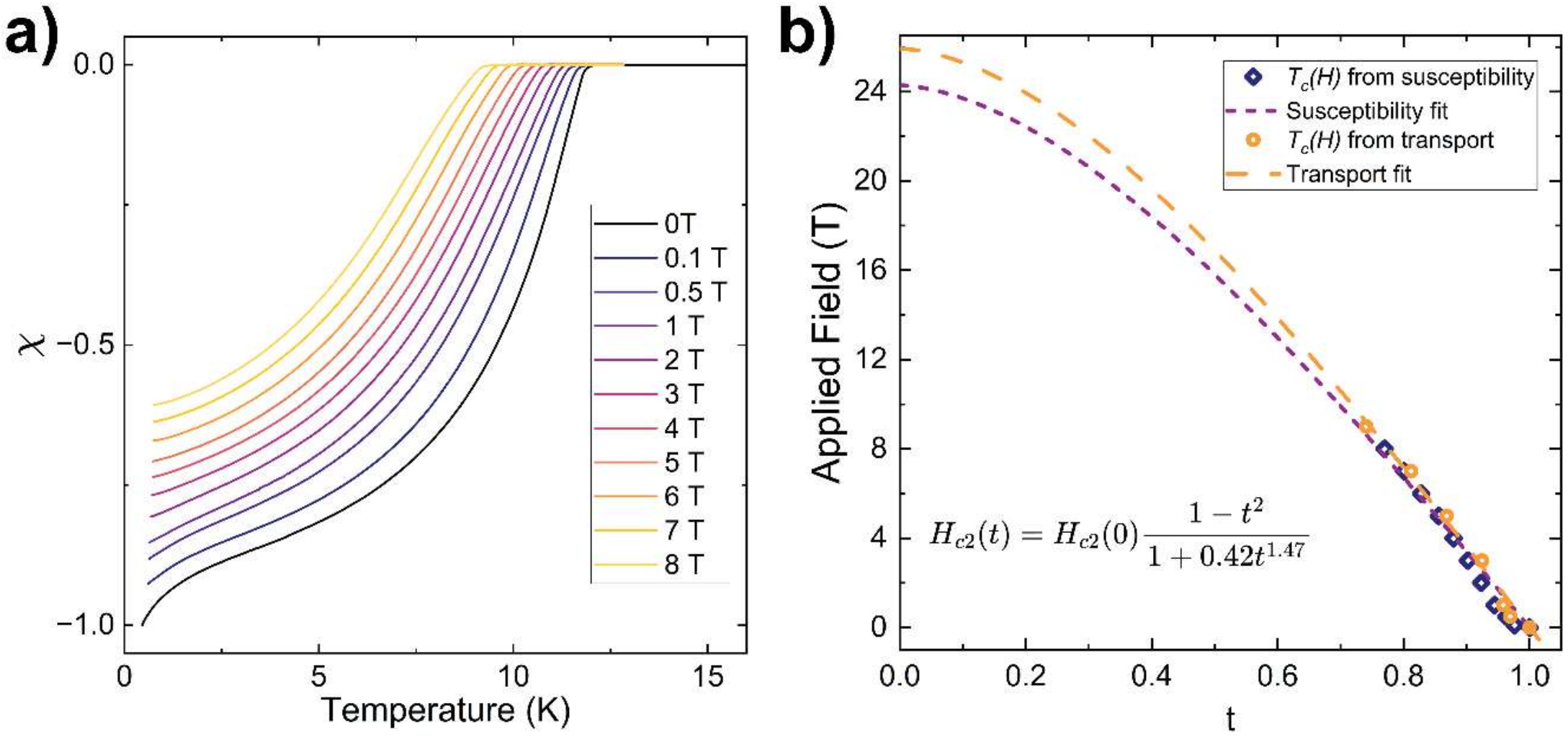

**Figure 9.** Change in normalized magnetic susceptibility with temperature (a) from tunnel diode resonator measurements in zero field and different field cooling. The change in critical temperature with applied field as measured in (a) is plotted in (b) in purple and the upper critical field is fit with the purple dashed line from the inset equation.[39] The orange data points and dashed lines correspond to the change in critical temperature with applied field from Figure 7(b), which is also fit to the inset equation. The variable t is defined as t=$T/T_c$ in (b).

## 3. Discussion

Given reported reduction times in the square-planar nickelates, which can easily reach 6 h for a film less than 20 nm thick, it is surprising that $LiNbO_3$ thicknesses exceeding 200 µm can be reduced to bulk $LiNbO_2$. The topotactic nickelate reduction is well-established as being dominated by diffusion kinetics, where a simple Fickian diffusion model suggests that the diffusion time, $\tau$, is related to the diffusion distance of oxygen, $L$, and the diffusion coefficient, $D$, *via*:

$$\tau \approx L^2/4D$$

In our $LiNbO_2$ system, a diffusion limited process across 200 µm should then require a treatment time approximately about $10^8$ times longer, assuming similar kinetics. Since a 200 µm layer of $LiNbO_3$ is reduced in 480 h, the increased temperature must increase the reaction rate significantly. The diffusion coefficient, $D$, is a function of the activation energy, $E_a$, and the temperature, $T$.

$$D = D_0 \exp(-\frac{E_a}{k_B T})$$

A typical infinite layer nickelate reduction is conducted at 300 °C and we conduct the $LiNbO_2$ reduction at 650 °C. For the sake of comparison, we can look at a ratio of diffusion coefficients for $LiNbO_2$ and the infinite layer nickelate films under the assumption that they have the same activation energy, while simultaneously considering the different reduction temperatures.

$$D_{Niobate}/D_{Nickelate} = \exp[-\frac{E_a}{k_B}\left(\frac{1}{T_{Nickelate}} - \frac{1}{T_{Niobate}}\right)]$$

However, most reasonable values for $E_a$ (in the 1.0 to 1.5 eV range) imply that either the nickelate or $LiNbO_2$ reduction would not finish in the observed time. Instead, it is likely that either the $LiNbO_3$ → $LiNbO_2$ reduction process has a lower activation energy, or the reduction is mediated through a different process which is not limited by oxygen diffusion. We speculate that the non-topotactic structural transformation creates a highly permeable product layer and a sharp advancing phase-formation front, shown in Figure 1, requiring oxygen to only diffuse to the grain boundaries which have much higher oxygen mobility.

Structural analysis from XRD and compositional analysis from SIMS and XPS together provides insight into the $LiNbO_2$ phase formation in the sample during $CaH_2$ reduction. We postulate that $LiNbO_2$ nucleates with the c-axis normal to the surface of the crystals, as the 001 basal plane is the lowest energy surface for materials in the $P6_3/mmc$ space group. The loosely packed and high surface energy Y-128° crystals help to improve the speed of nucleation and reconstruction at the surface. Oxygen is then easily reduced through grain boundaries between the nucleated $LiNbO_2$ as the phase front continues to advance and the $LiNbO_2$ crystals grow in a columnar fashion.

During the extreme high temperature reduction, some fraction of the lithium precipitates out of the $NbO_2$ matrix. This is explicitly demonstrated by both SIMS and XPS, which show a significant increase in Li content near the surface. However, XPS on the polished $LiNbO_2$ region also reveals that the vast majority of Li remains within the sample. This seeming contradiction may be resolved by examining lateral $6Li^+$ SIMS intensity maps (see supplemental Figure S3), which show a significant degree of lateral heterogeneity and clustering. We conclude that a significant fraction of the Li content precipitates out of the $LiNbO_2$ lattice, forming Li-rich regions, most likely near grain boundaries. Thus, while the average Li composition remains largely unchanged, significant lithium deficiency is present locally in the sample, inducing superconductivity in approximately 77% of the $LiNbO_2$ volume.

## 4. Conclusion

We successfully utilized $CaH_2$ reduction to induce a bulk, phase transformation from insulating $LiNbO_3$ single crystals to superconducting $LiNbO_2$. Cross-sectional SEM/EDS as well as powder XRD measurements confirm the bulk conversion from $R3c$ ($LiNbO_3$) to the $P6_3/mmc$ ($LiNbO_2$) phase. SIMS and XPS indicate that the requisite hole-doping to induce superconductivity is achieved during the $CaH_2$ reduction due to a combination of delithation and hydrogenation. By optimizing reduction temperature and time we achieve onset critical temperatures as high as 14.3 K. Low-temperature magnetization and radio-frequency magnetic susceptibility measurements confirm a bulk superconducting volume fraction of approximately ~77% and an upper critical field of approximately 26 T. These results show that $CaH_2$ reduction possesses the ability to facilitate bulk reconstructive phase transformations of complex metal oxides. Future studies should explore the application of extreme hydride reduction to discover superconducting phases in other highly distorted, non-perovskite metal oxides.

## 5. Methods

### 5.1 $CaH_2$ Reduction

The $LiNbO_3$ single crystals with a size of 10 mm x 10 mm x 0.5 mm were embedded with 0.5g finely ground $CaH_2$ powder and sealed in a quartz ampule under approximately 1E-3 Pa vacuum. The sample was heated at a rate of 10 °C/min and kept at 650 °C for different times, between 72 and 720 h. After the hydride treatment, the substrate was cleaned with a diluted $NH_4Cl$/methanol solution and blow dried by using the $N_2$ gas to remove excess $CaH_2$ and CaO left behind by the reaction. This treatment process is similar to what is reported by Yajima et al and Liu et al.[13, 15, 21]

### 5.2 Compositional and Structural Characterization

Samples were prepared for SEM by cutting with a diamond scribe and polishing the cross-section with incrementally increasing grit. The samples were then analyzed using a **Hitachi** SU-70 ultra high-resolution SEM equipped with a **Bruker** XFlash 6|60 EDS detector. EDS maps and line scans were acquired using **Bruker** Esprit 2.2 software. Time-of-flight secondary ion mass spectroscopy (ToF-SIMS) was performed using a 20 keV $Ar^{+}_{2600\pm1000}$ cluster beam to sputter into the sample, while analysis was performed using a 30 keV $Bi_3^+$ liquid metal ion source. The charge/mass ratios of sputtered ions were determined *via* a time-of-flight mass analyzer. We performed depth profiling by alternating the sputtering and analysis beams with 0.5 s of charge compensation per cycle. The analysis beam was rastered across an area of approximately a 500 μm × 500 μm area with a lateral resolution of (128 × 128) pixels, while the sputtering beam was confined inside that window to a region ranging from 300 μm × 300 μm to 400 μm × 400 μm, depending on the sample. The corresponding ion doses were $1.9 \times 10^9$ ions/cm$^2$ (0.12 pA) for $Bi_3^+$, and between $2.1 \times 10^{14}$ ions/cm$^2$ to $2.6 \times 10^{14}$ ions/cm$^2$ (5.1–6.4 nA) per cycle for the cluster source due to day-to-day fluctuations in the beam current. Residual chamber pressures were maintained below $5\times10^{-7}$ Pa to minimize the hydrogen background. Both positive and negative ions were measured, using separate sputtering craters in different parts of the sample. Spectra were analyzed using SurfaceLab to define the relevant region of interest, perform mass calibrations, identify peaks associated with the appropriate compounds, and to extract the total integrated peak intensity as a function of sputter time. A **Kratos** Axis UltraDLD spectrometer under a base pressure of 10-9 Torr (10-7 Pa) was used to acquire XPS spectra. A monochromatic Al Kα (10 mA, 15 kV) source was used, and the scans were carried out at 20 eV pass energy. A charge neutralizer was used for insulating $LiNbO_3$ with a filament current of 1.8 A and charge balance of 3.5 V but it was not used for $LiNbO_2$. XPS spectra were analyzed with Casa XPS.[31] The binding energy of all the spectra were referenced to hydrocarbon peak in C 1s spectra at 284.8 eV. The spectra for raw $LiNbO_3$ substrate were fitted with 70% Gaussian and 30% Lorentzian mixed peaks with a Shirley background. The Nb 3*d* peaks for the hydride reduced, $LiNbO_2$, samples were fit with Lorentzian Asymmetric line shapes, designated as LA(1.2, 5) in CasaXPS with a Shirley background. Li 1*s* and O 1*s* spectra were fit with 70% Gaussian and 30% Lorentzian mixed peaks for both the $LiNbO_3$ and $LiNbO_2$ samples. The Nb $3^+/5^+$ ratio was determined from the Nb 3*d* spectra. This ratio was fixed for the Nb 4*s* peaks to more accurately fit the overlapping Li 1*s* peak. Uncertainties reported in the chemical formula represent one standard deviation (k=1) derived from the XPS atomic concentration quantification. The goodness of fit was evaluated by minimizing the statistical residuals, while fitting uncertainties were constrained and validated using prior chemical and structural knowledge of the sample. The reduced $LiNbO_2$ region of samples were crushed, and the powder XRD patterns were measured using a **Bruker** D8 Powder Diffractometer (CuKα λ = 1.54 Å). Reitveld Refinement of the powder patterns was performed using the GSAS-II software.[42] As reduced crystals were measured in **Photonic** Laue Back scattering system to determine orientation of the crystals (CuKα λ = 1.54 Å). A **Rigaku** SmartLab II benchtop HRXRD (CuKα λ = 1.54 Å) with a Hypix-4000 detector and a Ge (220) 2-bounce monochromator was used to collect the 2D HRXRD scans. 2D HRXRD scans were integrated into one dimension using the **Rigaku** SmartLab Studio II software.

### 5.3 Superconducting Property Characterization

The low temperature transport was carried out in a **Quantum Design** 9 Tesla PPMS by using the linear four-probe method from 300 K down to 1.8 K. Low temperature magnetization measurements were carried out in a 7 Tesla **Quantum Design** MPMS-XL system. ZFC and FC measurements were taken from 300 K down to 1.8 K. ZFC measurements were carried out with 1 mT applied field. A self-oscillating tunnel diode resonator with operating resonance frequency of around 14 MHz was used to study magnetic susceptibility. The magnitude of the ac excitation magnetic field on the sample inside the inductor is roughly 2 μT well below the lower critical field. The temperature dependent normalized magnetic susceptibility $\chi$ was extracted from the resonance frequency shift *df* using the relation $\chi = \frac{\mathrm{df}}{f_o} - 1$ where $f_o$ is the resonance frequency without the sample. The FC curves were obtained by applying DC magnetic fields using the 9 Tesla superconducting magnet. Detailed description about the technique can be found elsewhere.[41]

### Acknowledgements

This work was funded by the Breakthrough Energy Fellow program. Ryan Paxson was supported by the Laboratory for Physical Science Qubit Collaboratory. The authors acknowledge the Maryland Quantum Materials Center. The project was partly supported by the Air Force Office of Scientific Research Grant No. FA9950-22-1-0023, the National Science Foundation DMREF program (DMR2523217), and the NIST Center for Neutron Research. A part of the research was sponsored by the Army Research Laboratory and was accomplished under Cooperative Agreement Number **W911NF-25-2-0084**. The views and conclusions contained in this document are those of the authors and should not be interpreted as representing the official policies, either expressed or implied, of the Army Research Laboratory or the U.S. Government. The U.S. Government is authorized to reproduce and distribute reprints for Government purposes notwithstanding any copyright notation herein. Unless otherwise noted, the NIST work was funded solely by the US government. Certain commercial equipment, instruments, or materials, commercial or non-commercial, are identified in this paper in order to specify the experimental procedure adequately. Such identification does not imply recommendation or endorsement of any product or service by NIST, nor does it imply that the materials or equipment identified are necessarily the best available for the purpose. Partial support of AJG was provided by the Center for High Resolution Neutron Scattering, a partnership between the National Institute of Standards and Technology and the National Science Foundation under Agreement No. **DMR-2010792**. We acknowledge Heshan Yu for his work in the preliminary stage of the project. Work at Ames National Laboratory was supported by the U.S. Department of Energy (DOE), Office of Science, Basic Energy Sciences, Materials Science and Engineering Division. Ames Laboratory is operated for the U.S. DOE by Iowa State University under contract **DE-AC02-07CH11358**.

**Data Availability Statement**

The data that support the findings of this study are available from the corresponding author upon reasonable request.

# Supporting Information

**Synthesis of Bulk Superconducting $LiNbO_2$ Crystals through $CaH_2$ Reduction**

*Ryan Paxson[1,2,]*, Stephanie J. Hong[3], Bicky Moirangthem[6], Parham Kabirifar[5], Saya Takeuchi[1], Tianyu Li[3], Chih-Yu Lee[1], Haotong Liang[1], Kamal Joshi[6], Amlan Datta[6], Makariy Tanatar[6], Shanta Saha[2], Peter Zavalij,[3] Ruslan Prozorov[6], Alexander J. Grutter[4], Efrain E. Rodriguez[1,2,3,]*, and Ichiro Takeuchi[1,2,]**

*[1]Department of Materials Science and Engineering, University of Maryland, College Park, MD 20742*
*[2]Maryland Quantum Materials Center, University of Maryland, College Park, MD 20742*
*[3]Department of Chemistry & Biochemistry, University of Maryland, College Park, MD 20742*
*[4]NIST Center for Neutron Research, National Institute of Standards and Technology, Gaithersburg, MD 20899*
*[5]Advanced Imaging and Microscopy (AIM) Lab, Maryland Nanocenter, University of Maryland, College Park, MD 20742*
*[6]Ames Laboratory and Department of Physics and Astronomy, Iowa State University, Ames, IA 50011, USA*

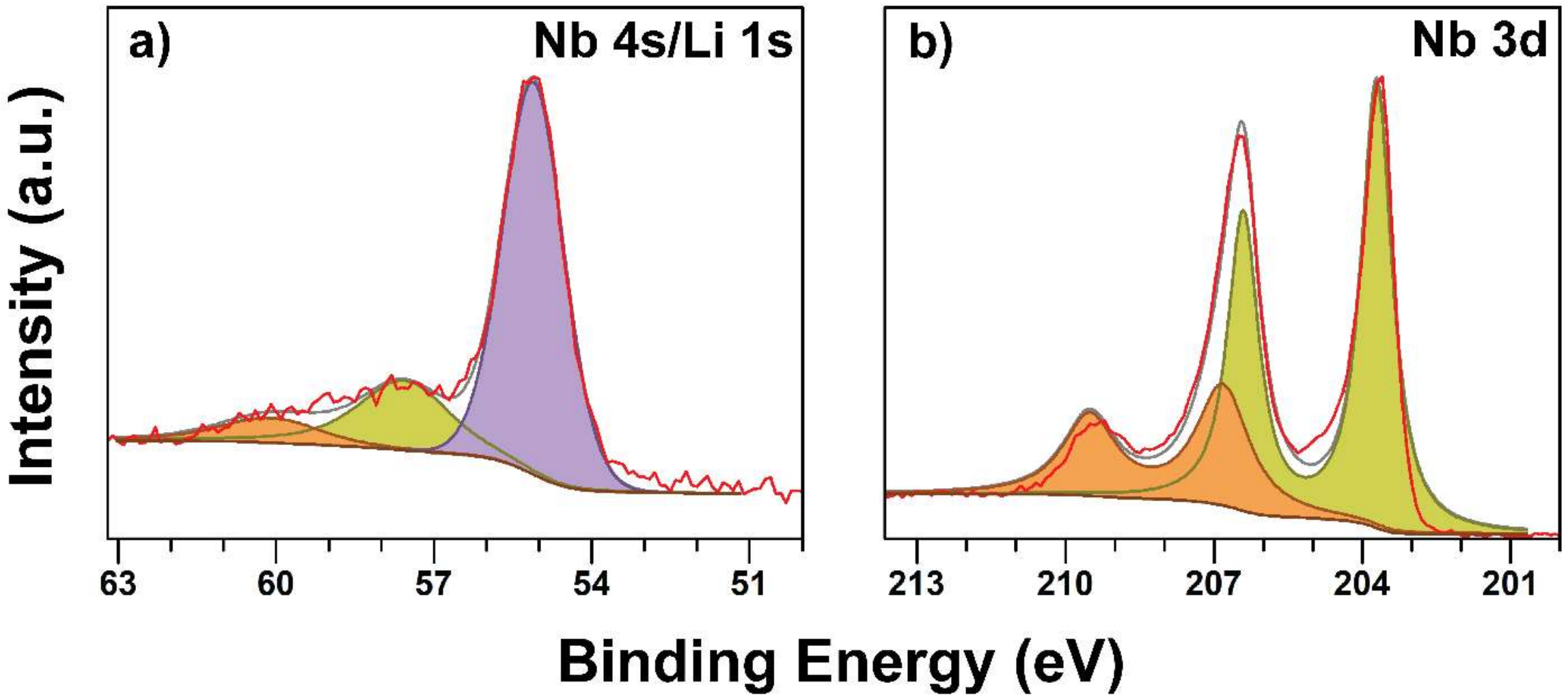

Figure S1. XPS spectra of as reduced $LiNbO_2$ sample. The Li 1*s* peak of the as reduced sample, (a) is very prominent and the calculated Li:Nb ratio is approximately 7:3. This increased presence of Li at the surface of the as reduced sample agrees with SIMS measurements shown in Figure 3(a). The Nb 3*d* spectra (b) seems relatively unperturbed from the excess Li at the surface. This lends to the idea that some of the Li form lithium oxides and oxyhydrides on the surface as it migrates there during the $CaH_2$ reduction.

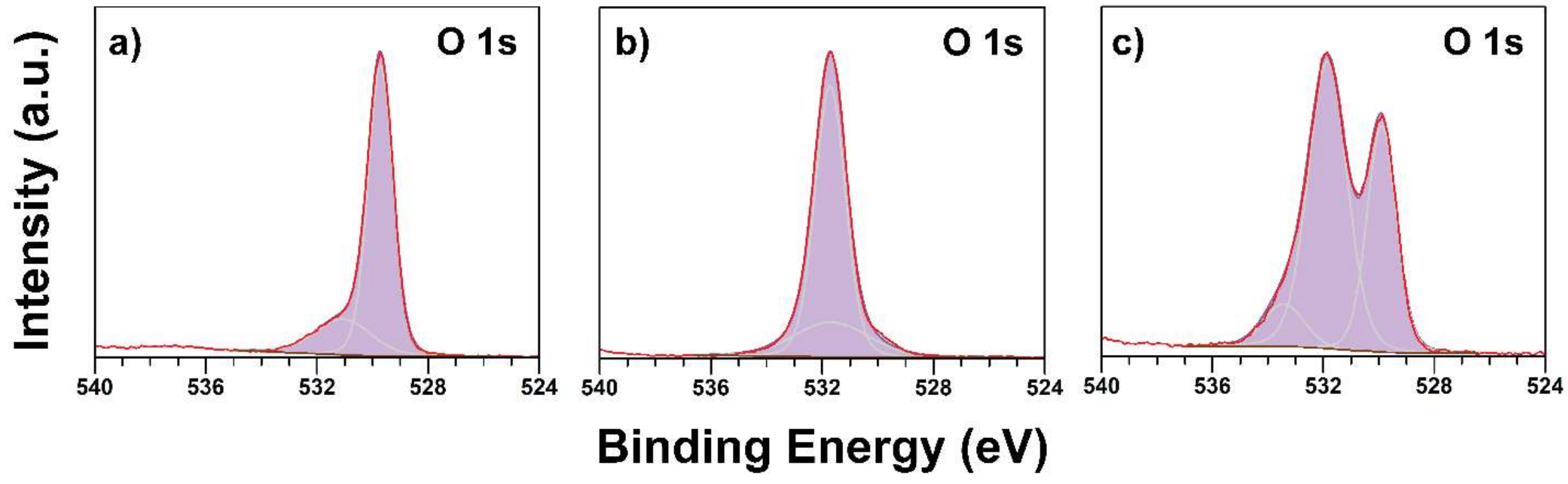

Figure S2. O 1*s* XPS spectra of control $LiNbO_3$ (a), as reduced $LiNbO_2$ (b), and polished $LiNbO_2$ (c).

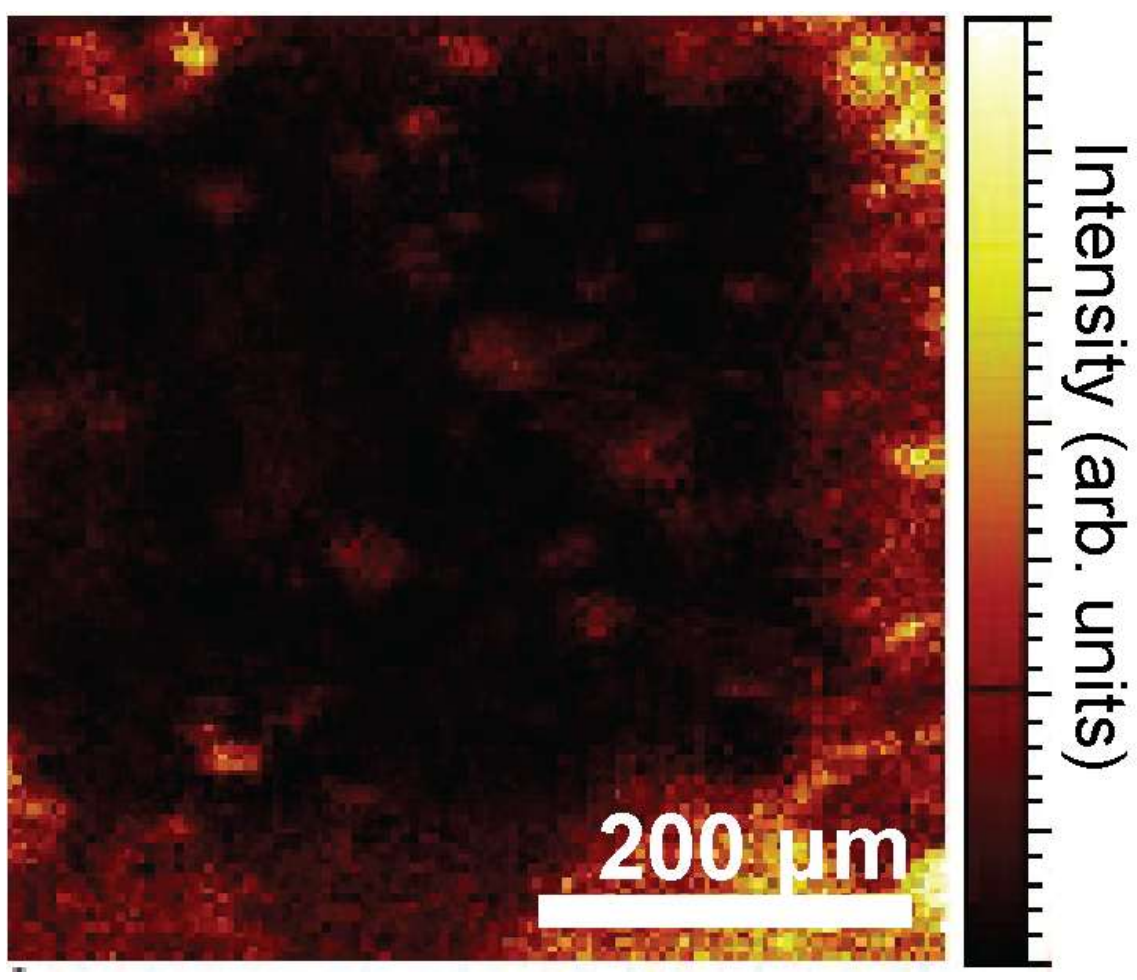

Figure S3. SIMS area mapping of $^{6}Li^{+}$ ion distribution showing lithium clustering on the surface and at depth into the sample. Brightness corresponds to relative secondary ion intensity.